\documentclass[pre,twocolumn,showpacs]{revtex4}
\usepackage{epsfig}
\usepackage[nooneline]{subfigure}
\usepackage[]{natbib}
\begin{document}

\title{ 
Time averaged Einstein relation and fluctuating diffusivities
for the L\'evy walk}

\author{D. Froemberg}
\author{E. Barkai}
\affiliation{Department of Physics, Institute of Nanotechnology and Advanced Materials, Bar Ilan University, Ramat-Gan
52900, Israel}

\pacs{02.50.-r, 05.40.Fb}

\begin{abstract}
The L\'evy walk model is a stochastic framework of enhanced diffusion
with  many applications in physics and biology.
Here we investigate the time averaged mean squared displacement $\overline{\delta^2}$  often used
to analyze single particle tracking experiments.
The ballistic
phase of the motion is non-ergodic and we obtain analytical expressions
for the fluctuations of $\overline{\delta^2}$.
For enhanced sub-ballistic diffusion we observe
numerically apparent ergodicity breaking on long time scales.
As observed by Akimoto \textit{Phys. Rev. Lett.}
\textbf{108}, 164101 (2012)
deviations of temporal averages  $\overline{\delta^2}$
from the ensemble average $\langle x^2 \rangle$
depend on the initial preparation of the system, and here
we quantify this discrepancy from normal diffusive behavior.
Time averaged response to a bias is considered and the resultant 
generalized Einstein relations are discussed.
\end{abstract}

\maketitle

In recent years  there has been  growing interest in the physics of 
weak ergodicity breaking \cite{B92,Bouchaud,Aaronson}. In statistical mechanics
the ergodic hypothesis states
that ensemble averages and time averages are equal in the limit of
long measurement times. Weak ergodicity breaking is found in systems
whose temporal dynamics is governed by broad
power law distributed waiting times 
with a diverging mean \cite{B92,Bouchaud}.
Weak ergodicity breaking allows for the exploration of the whole phase
space (unlike strong ergodicity breaking) yet ergodicity is not
attained since the diverging time scale of the dynamics
always exceeds the measurement time
\cite{Barkai1}. The lack of a time scale for the
dynamics leads to distributional
limit theorems for time averaged observables which are not trivial
\cite{Aaronson,Golan,RB07,HeBur08,Lub08}
while in the ergodic phase the distribution of the time averages
are delta functions centered on the ensemble averages and in that
sense are trivial. Weak ergodicity breaking is observed
in many systems ranging from blinking quantum dots \cite{Dahan,P1day}
 (where sojourn times
in on and off states are power law distributed) to models of glassy dynamics
\cite{B92},
and diffusion of molecules in the cell environment \cite{Krapf,Jeon,P2day}. 

An observable that was extensively studied is the mean square displacement
(MSD) of a diffusing particle \cite{P2day}. 
Let us first define the time averages.
Experimentalists routinely track individual trajectories of particles
and use the information
for precise measurements of diffusion constants. 
The time averaged MSD  
is defined through the path $x(t)$ in terms of 
\begin{equation}
\overline{\delta^2} = {1 \over T - \Delta} \int_0 ^{T- \Delta} [ x(t+\Delta) - x(t)]^2 {\mathrm d } t
\label{eq01}
\end{equation}
with the lag time $\Delta$ much smaller than the measurement time
$T$. 
In the case of Brownian motion, due to the stationary increments 
$\lim_{T \to \infty} \overline{\delta^2} = 2 D \Delta$
is precisely the same as the MSD averaged over a large
ensemble of  particles
$\langle x^2 \rangle = 2 D \Delta$, indicating
ergodicity in the MSD sense. Here $D$ is the diffusion constant.
Now consider a weak force $F$ acting  on the 
particle,
which will induce a net drift $\langle x \rangle_F \propto \Delta$.
All along this work $\langle \cdots \rangle$ denotes ensemble averages
while $\overline{\cdots}$ stands for time averaging. 
For Brownian particles the response to the field is related to the fluctuations 
via the celebrated Einstein relation
$\langle x \rangle_F = F \langle x^2 \rangle / (2 {\cal T})$ \cite{Bouchaud,BarFleu98} where ${\cal T}$ is the temperature 
and all along this work $k_B =1$.
Since the Brownian process is ergodic 
the Einstein  relation will hold also  for the time averaged response defined
according to
\begin{equation}
\overline{\delta}_F = \frac{1}{T-\Delta}\int_0^{t-\Delta}  \left[  x(t+\Delta) - x(t) \right]  dt,   \label{EATAdrift}
\end{equation}
thus $\overline{\delta}_F = F \overline{\delta^2} / (2 {\cal T})$. 

Very recently Akimoto 
investigated temporal averages
of anomalous diffusion and response to bias  
within the framework of deterministically generated L\'evy walks 
\cite{Akimoto12}. In this well investigated and widely applicable  process
the diffusion is anomalous. Here two interesting issues arise.
The first is the question of ergodicity of these processes,
and the second the applicability of the Einstein relation
to the time averages. 
Though time averaged response to a bias was found to be  intrinsically
random   surprisingly the 
``
temporal averaged MSDs  are not random'' \cite{Akimoto12}. 
This is diametrically
opposed to 
results  in previous examples of weak ergodicity breaking where
temporal averages are intrinsically random
\cite{Golan,RB07,HeBur08,MarBar05}.  
For that reason we  analytically investigate the previously
ignored  non-trivial
fluctuations of the time averaged MSD showing that the fluctuations
are universal. 
We then formulate a new Einstein relation for the time
averages. We show that the Einstein relation for time averages
differs considerably from the corresponding Einstein relation for
the ensemble averages.

The L\'evy walk model is a generalization of 
the classical Drude model describing 
a particle moving with constant velocity and
changing its direction randomly. While in the Drude model exponential 
waiting times between turning events due to strong collisions result in a Markov process,
the L\'evy walk model  postulates power law distributed waiting times between
randomization events resulting in long flights \cite{ShleWest87,P2dayBBM,Mag12}. 
The L\'evy walk \cite{ShleWest87}
describes enhanced transport phenomena 
in many systems, ranging from
chaotic diffusion to animal foraging patterns
\cite{GeiZach85,ZumKlaf93,SolSwin93,GeiZach88,MargBarJCP04,MarBar05,Humph10,Gal10}.
For some very recent applications see also \cite{ZabDe11,DhaSai12,Sagi,KessBar12}.
The ubiquity of L\'evy walks makes it particularly interesting to
characterize and quantify their ergodic properties leading to a better understanding
of the physics at the core of such processes. 

\textit{L\'evy walk:  model and ensemble averaged MSD.}
Super-diffusion based on power law waiting times 
 is naturally described by the L\'evy walk model \cite{P2dayBBM}. 
We consider a particle alternating its velocity
between $+v_0$ and $-v_0$ at random times. The times $0<\tau<\infty$
between turning events
are independent, identically distributed random variables
with a common  probability density function (PDF) $\psi(\tau)$.
The position of the particle is $x= \int_0 ^t v(t^\prime) {\mathrm d} t^\prime$,
so that the particle starts at $t=0$ with velocity $+v_0$, travels a distance
$v_0 \tau_1$ with $\tau_1$ drawn from $\psi(\tau)$, and after that is 
displaced $-v_0 \tau_2$. The process is then renewed.
The PDF of flight times $\tau$ 
is power law distributed, $\psi(\tau) \sim
A \tau^{-(1 + \alpha)}/|\Gamma(-\alpha)|$.
When $0<\alpha<1$ the mean $\langle \tau \rangle$ diverges,
while for $1<\alpha<2$ it is finite though $\langle \tau^2 \rangle=\infty$. 
Our working example in simulations will  be $\psi(\tau) = \alpha \tau^{-(1+\alpha)}$ for $\tau>1$.
Importantly, the displacements $\pm v_0 \tau$ are broadly distributed 
though they never become larger than $\pm v_0 t$. 

\begin{figure}[hbt]
\centering{
{\includegraphics[width=.235\textwidth]{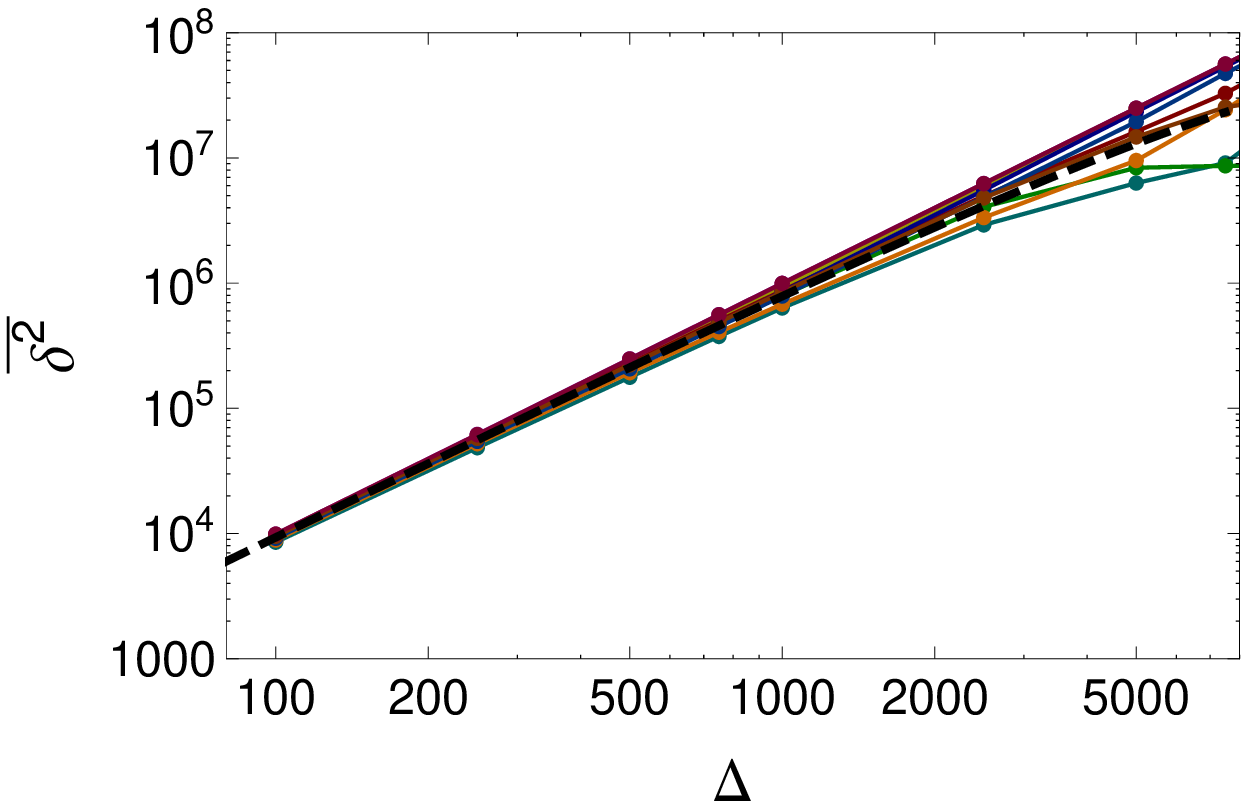}}
\hspace{.01cm}
{\includegraphics[width=.235\textwidth]{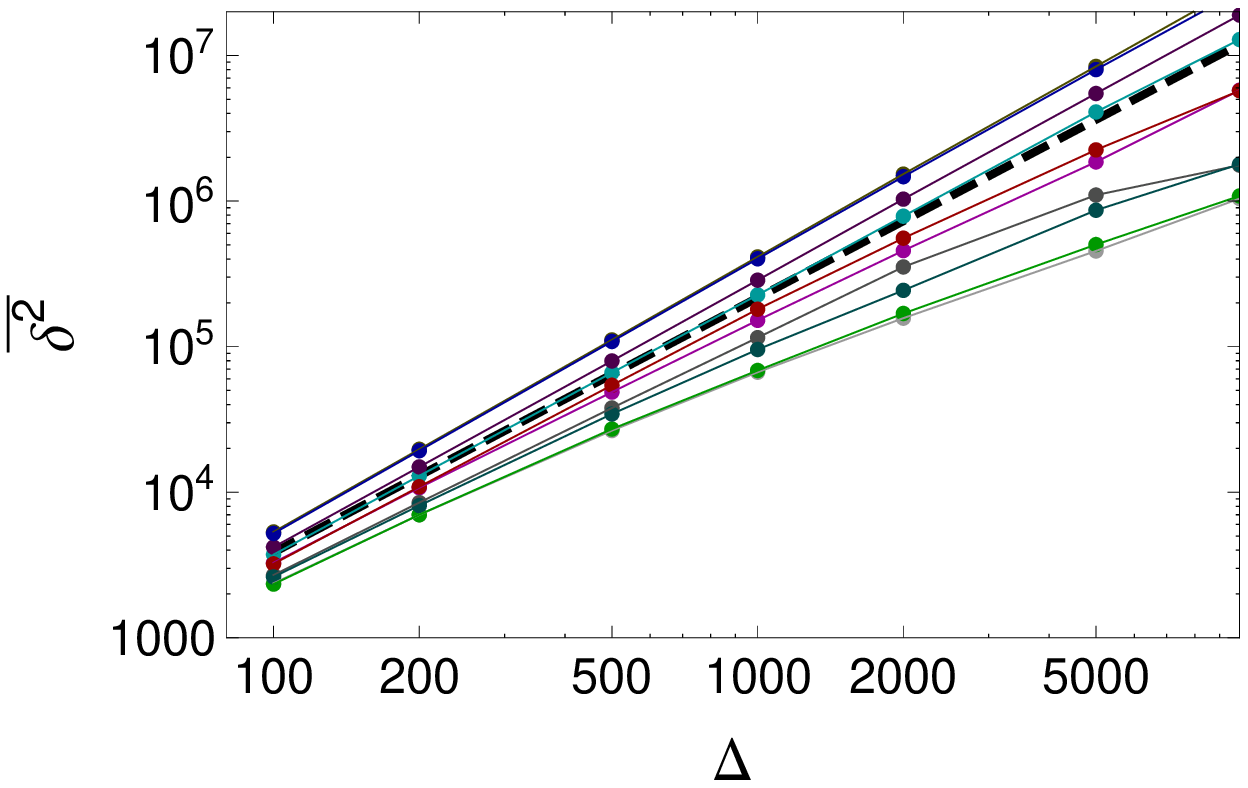}}
\caption{\label{fig1} $\overline{\delta^2}$ of ten sample trajectories vs. lag time $\Delta$, 
$T=10^{4}$, $\alpha=1/2$ (left); $T=10^5$, $\alpha = 5/4$, dashed lines indicate ensemble averages $\langle\overline{\delta^2}\rangle$ Eqs. (\ref{eq08}) and (\ref{taDel}).
Larger times $T$ and smaller lag times $\Delta$ result in smaller fluctuations. For comparable times $T$, the 
fluctuations in the enhanced case are larger than in the ballistic case.}}
\end{figure}

The ensemble averaged
MSD is \cite{Masoliver, ZumKlaf90, P2dayBBM}
\begin{equation}
\langle x^2 \rangle \sim  \left\{
\begin{array}{c c}
v_0^2 (1 - \alpha) t^2 & 0<\alpha<1 \\
2 K_\alpha t^{3-\alpha}  & 1<\alpha<2 .
\end{array}
\right.
\label{eq02}
\end{equation}
The case $0<\alpha<1$ is called the ballistic phase, while
we refer to the parameter range
$1<\alpha<2$ as enhanced diffusion which is sub-ballistic. 
Here the anomalous diffusion coefficient is given by
\begin{equation}
K_{\alpha} = v_0^2 {A (\alpha -1) \over \langle \tau \rangle \Gamma\left(4 -\alpha\right)} .
\label{eq03}
\end{equation}
Note that  this transport coefficient was derived for a process which started
at time $t=0$. 
For both ballistic and enhanced regimes simulations reveal fluctuations of $\overline{\delta^2}$ at finite $T$, 
which surprisingly are more pronounced in the latter case (Fig. \ref{fig1}). 
Before we turn to these fluctuations, we will study
the ensemble averages $\langle\overline{\delta^2}\rangle$, i.e. the mean of 
the distributions of the $\overline{\delta^2}$, denoted by the dashed lines in Fig. \ref{fig1}.

Averaging Eq. 
(\ref{eq01}) we notice a relation between 
$\langle \overline{\delta^2} \rangle$
and the ensemble averaged position correlation function
\begin{equation}
\langle \overline{\delta^2} \rangle=
 \int_0 ^{T-\Delta} {
\langle x^2(t+\Delta) \rangle + \langle x^2 (t) \rangle - 2 \langle x(t) x(t+\Delta)\rangle \over T-\Delta} {\mathrm d} t
\label{eq04}
\end{equation}
The correlation function $\langle x(t_1) x(t_2) \rangle$ is related to the 
velocity correlation function as
\begin{eqnarray}
\langle x(t_1) x(t_2) \rangle &=& \int_0 ^{t_1} {\mathrm d} t^\prime_1 \int_{0} ^{t_2} {\mathrm d} t^\prime_2 \langle v(t^\prime_1) v(t^\prime_2) \rangle .\label{eq05}
\end{eqnarray}
Since $v(t_1) = v(t_2)$ ( or $v(t_1) = - v(t_2)$) when the number
of transitions $n$ in the time interval 
$(t_1,t_2)$ is even (or odd),
we have
\begin{equation}
\langle v(t_1) v(t_2) \rangle = \sum_{n=0} ^\infty (-1)^n v_0^2 p_n(t_1,t_2),
\label{eq05a}
\end{equation}
where $p_n(t_1,t_2)$ is the probability for $n$ transitions of direction
in the time interval $(t_1,t_2)$.
The behavior of the velocity correlation function Eq.
(\ref{eq05a})  was studied by Godr\`eche and Luck \cite{GodrLuck01}
and it is  described
by two limits depending on the value of $\alpha$. 

\textit{The ballistic phase.}
For $\alpha<1$  the dynamics is free of a time scale since $\langle \tau \rangle = \infty$  so that
the particle will get stuck in a velocity state (either
$+v_0$ or $-v_0$) for a duration of the  order of the measurement time.
Hence the dominating term in
Eq.
(\ref{eq05a})
is $n=0$ and only the persistence probability $p_0(t_1,t_2)$ 
is important in the scaling limit of the problem \cite{GodrLuck01}, 
\begin{equation}
\langle v(t_1) v(t_2) \rangle \simeq v_0^2 p_0(t_1,t_2)=
v_0 ^2 {\sin \pi \alpha \over \pi} B({t_1 \over t_2};\alpha,1-\alpha) ,
\label{eq06}
\end{equation}
where $B(z;a,b)$ is the incomplete Beta function and $t_2\geq t_1$. 
This velocity correlation function cannot be expressed as a function of the
time difference $|t_2-t_1|$, reflecting the non-stationarity of the process. 
Inserting Eq. (\ref{eq06}) in Eq. (\ref{eq05})  and integrating
we get 
\begin{widetext}
\begin{equation}
\langle x(t_1) x(t_2) \rangle = v_0^2 {\sin \pi \alpha \over \pi} \left[ t_1 t_2 B \left( {t_1 \over t_2}; \alpha , 1- \alpha \right) -
{1 \over 2} (t_2)^2 B \left( {t_1 \over t_2} ; 1+\alpha,1-\alpha\right)
-{1 \over 2} (t_1)^2 B\left( {t_1 \over t_2}, -1+\alpha,1-\alpha\right) \right] - \alpha { (v_0 t_1)^2 \over 2} 
\label{eq07}
\end{equation}
\end{widetext}
which reduces to the first line of Eq. (\ref{eq02}) when $t_1=t_2$. 
In the limit $\alpha \to 0$ the particle remains in state
$+v_0$ or $-v_0$ for the whole duration of measurement time,
hence we expect and indeed get $\langle x(t_2) x(t_1) \rangle=
v_0^2 t_2 t_1$ which describes a deterministic motion. 
In contrast, for Brownian motion we have
$\langle x(t_1) x(t_2) \rangle = 2 D \mbox{min}(t_1,t_2)$
reflecting independent increments of the process.  
Compared with the diffusive case the L\'evy walk
exhibits strong correlations due to the long sticking times in the positive or negative
velocity states. For $\Delta\ll t_1$ we find
$\langle x(t_1) x(t_1 + \Delta) \rangle \sim \langle x^2(t_1) \rangle [ 1 + (\Delta/t_1)]$ thus 
the correlations are strong in the sense that they are increasing with $\Delta$. 

Inserting the  correlation function Eq. (\ref{eq07}) 
in
Eq. (\ref{eq04}) and integrating we find 
in the limit $\Delta/T << 1$
\begin{equation}
\langle \overline{\delta^2} \rangle \sim  v_0^2\left[ \Delta^2 -  { \sin \pi \alpha \over \pi \alpha} { 2 \Delta^2 \left( { \Delta \over T} \right)^{1-\alpha} \over 6 - 11 \alpha + 6 \alpha^2 - \alpha^3 } \right] .
\label{eq08}
\end{equation}
The leading term $\langle \overline{\delta^2} \rangle \sim (v_0 \Delta)^2$
was found in \cite{Akimoto12} and
corresponds to a deterministic, ballistic motion
with velocity $v_0$.
To see this insert
$[x(t+\Delta) - x(t)]^2 = (v_0 \Delta)^2$ in 
Eq. (\ref{eq01}) which yields $\overline{\delta^2}= (v_0 \Delta)^2$.

\begin{figure}[ht]
\centering{
{\includegraphics[width=.45\textwidth]{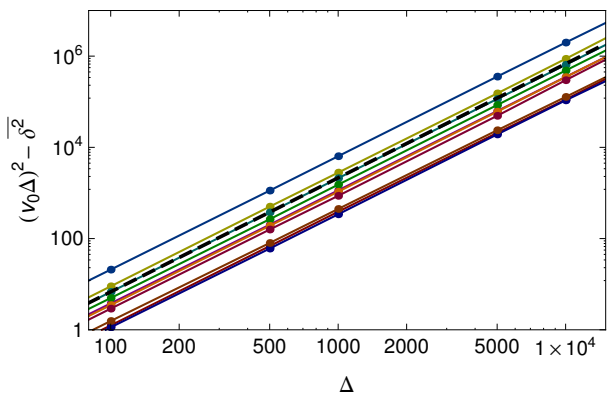}}
}
\caption{\label{fig2} Deviations from ballistic motion of the time averaged MSDs 
$( v_0 \Delta)^2 - \overline{\delta^2}$ versus the lag time $\Delta$ for ten
trajectories; $\alpha=1/2$, $v_0=1$, $T=10^8$. The dashed line denotes the theoretical ensemble average
Eq. (\ref{eq08})}
\end{figure}

More important 
are the fluctuations of the time averaged  MSD which 
quantify the ergodicity breaking.
To explore this issue note that a
particle not changing its direction at all $\overline{\delta^2} = (v_0 \Delta)^2$ 
corresponds to a ballistic path. If the particle changes its velocity
only once in the interval $(0,T)$, it is easy to show \cite{remark1} that 
$\overline{\delta^2} = (v_0 \Delta)^2 - (2/3) v_0 ^2 \Delta^3 / T$ for $T\gg\Delta$.
Thus a single switching event reduces $\overline{\delta^2}$ by a term $\chi^2$
proportional to $v_0 ^2 \Delta^3 / T$. 
If we have two transitions between $+v_0$ and
$-v_0$ states, the correction term is twice as large \cite{remark6}.
Altogether we deduce that for a random amount $n_T$ of switching events within 
the observation time $T$ 
\begin{equation}
\overline{\delta^2} = (v_0 \Delta)^2 -  \chi^2 n_T. 
\label{eq09}
\end{equation}
Notice that this result is valid for a single trajectory,
both $\overline{\delta^2}$ and $n_T$ are random.  
Once we find $\chi^2$ 
this equation gives the sought after fluctuations of the time averaged
MSD as will become clear soon.
Further, we see that the natural random variable is the shifted MSD
$(v_0 \Delta)^2 - \overline{\delta^2}$
which is plotted in Fig. \ref{fig2} versus the lag time
$\Delta$. Now the fluctuations are clearly visible unlike
the presentation in Fig. \ref{fig1}.

We now determine  $\chi^2$.
From renewal theory \cite{Feller, Bouchaud}  
the average number of switchings (renewals) is
$\langle n_T \rangle \sim T^\alpha/A\Gamma(1+\alpha)$.
Comparison of
the average of  Eq. (\ref{eq09}) with Eq. (\ref{eq08})
thus yields
\begin{equation}
\chi^2= \frac{2 \sin \pi\alpha A \Gamma(1+\alpha)}{\pi\alpha (6-11\alpha+6\alpha^2-\alpha^3)} 
{{v_0^2 \Delta^{3-\alpha}} \over T},
\label{eq10}
\end{equation}
a result valid for $\Delta\ll T$ \cite{remark2}.

\begin{figure}[hbt]
\centering{
{\includegraphics[width=.235\textwidth]{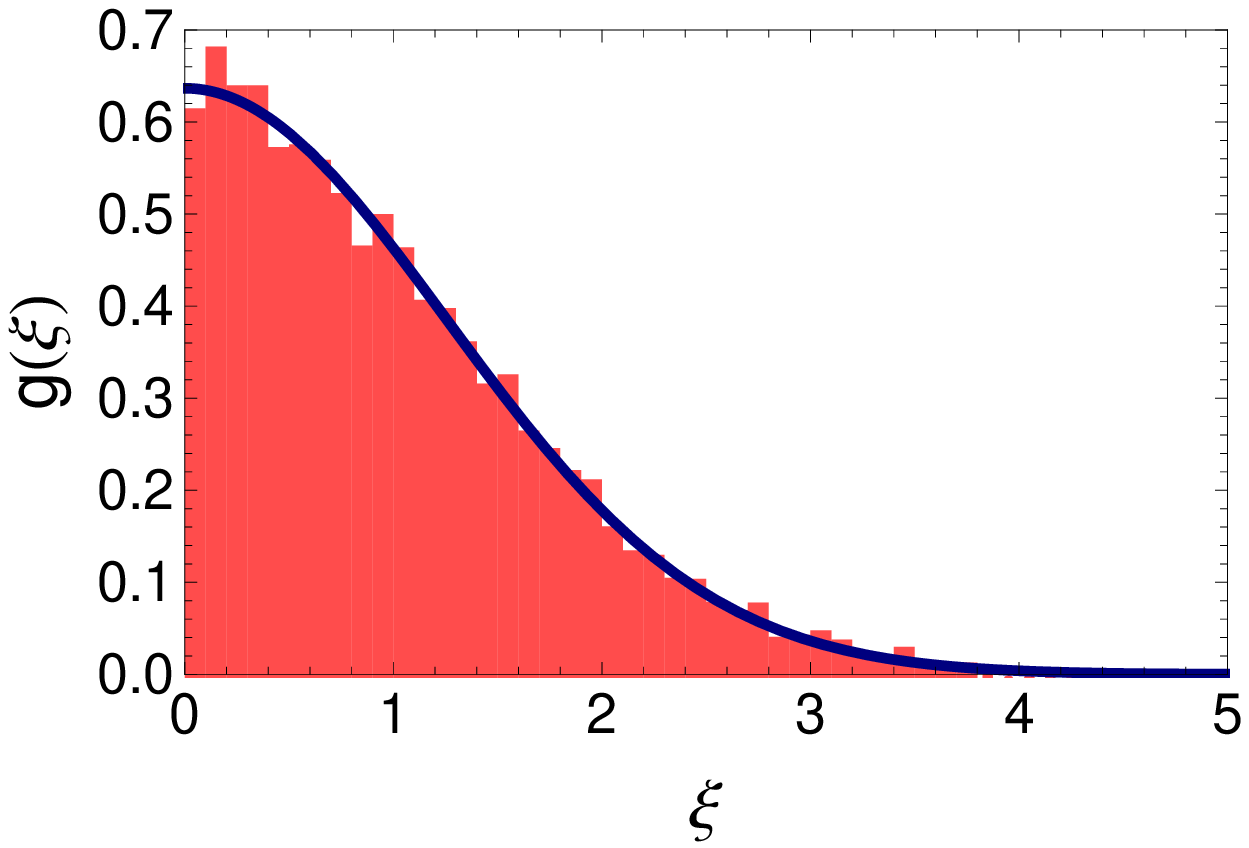}}
\hspace{.01cm}
{\includegraphics[width=.235\textwidth]{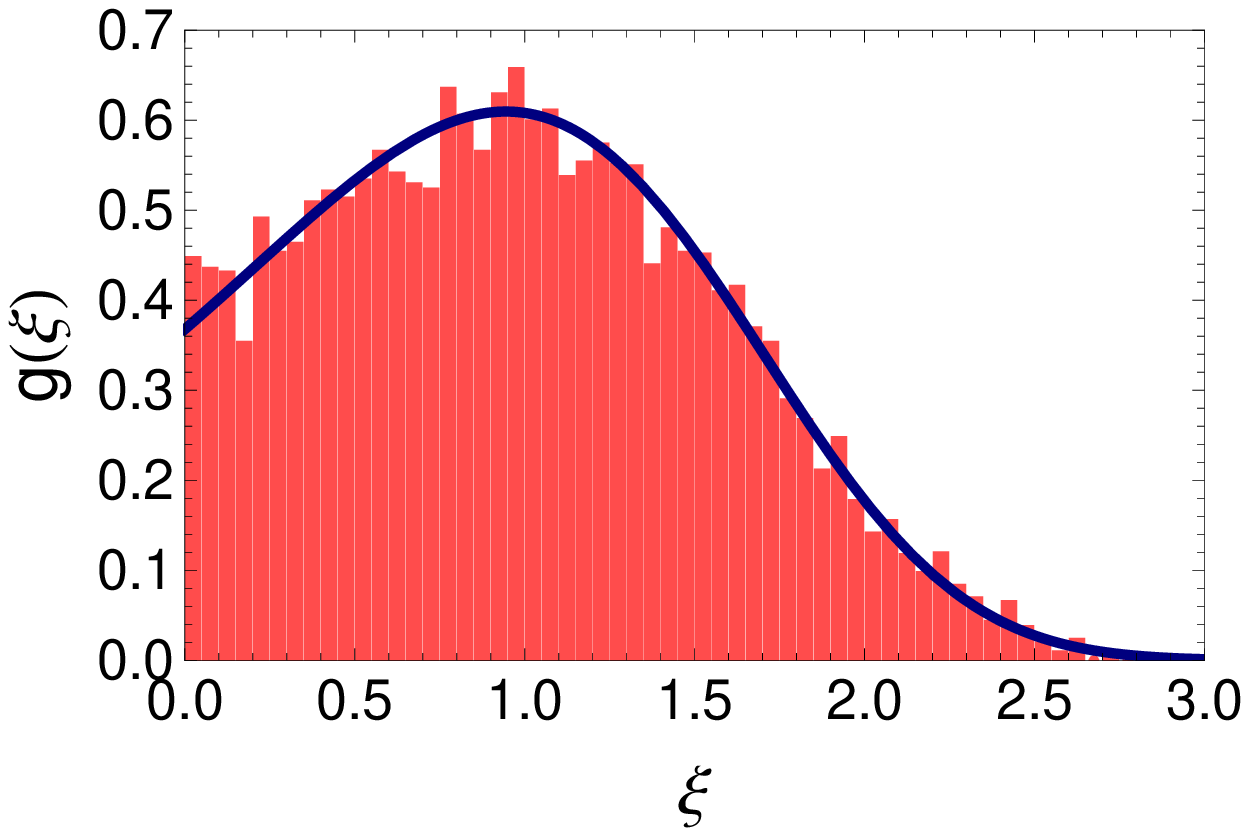}}}
\caption{\label{fig3} PDF of 
$\xi = ((v_0 \Delta)^2 - \overline{\delta^2}) / ((v_0 \Delta)^2 - \langle \overline{\delta^2} \rangle) $  
for $\alpha=0.5$ (left) and $\alpha=0.7$ (right).
Simulations (histograms) comply with theory (solid lines); $T=10^6$, $\Delta=1$. When $\alpha=0.5$ the peak 
of the PDF is on $\xi=0$; for $\alpha \to 1$ the peak tends to $\xi=1$. In that limit the fluctuations of $\xi$ vanish.}
\end{figure}

To quantify the fluctuations we introduce the dimensionless
random parameter
\begin{equation}
\xi = {   \overline{\delta^2}- (v_0 \Delta)^2\over   \langle \overline{\delta^2} \rangle- (v_0 \Delta)^2} = { n_T \over \langle n_T \rangle} 
\label{eq11}
\end{equation}
which has mean equal one. 
The fluctuations of the number of switchings $n_T$ are well known from
renewal theory \cite{Aaronson,Feller,Bouchaud} as they are determined by
the waiting time distribution $\psi(\tau)$ only.
The case $\alpha<1$ implies that L\'evy's central limit theorem holds
which gives
the PDF of $\xi$ 
\begin{equation}
g ( \xi) = { \Gamma^{1/\alpha} (1 + \alpha) \over \alpha \xi^{1 + 1/\alpha} }
l_{\alpha,1} \left[ { \Gamma^{1/\alpha} (1 + \alpha) \over \xi^{1/\alpha} } \right]. 
\label{eq12}
\end{equation}
Here $l_{\alpha,1}(t)$ denotes the one sided L\'evy stable PDF whose
Laplace transform is given by $\exp(- u^\alpha)$ \cite{Feller,Bouchaud,remark7}.
Fig. \ref{fig3} shows  excellent agreement between
the PDF Eq. (\ref{eq12}) and the respective simulation results. 
As mentioned in the introduction \cite{Akimoto12} showed that
transport i.e. time averaged response to external bias is random.
Eq. (\ref{eq12})  shows that also the time average diffusivity of the process 
is random though one must consider the shifted MSD defined in Eq. 
(\ref{eq11}) to observe the fluctuations typical for 
weak ergodicity breaking.

\textit{The enhanced diffusion phase.}
For $1<\alpha<2$ the dynamics has a finite time scale $\langle \tau \rangle$ 
and therefore one may naively expect the normal behavior
$\langle \overline{\delta^2} \rangle = \langle x^2 \rangle$. Similarly
to the ballistic phase we find the correlation function
\cite{remark3}
\begin{eqnarray}
&&\langle x(t_1)x(t_1+\Delta)\rangle =
\frac{K_{\alpha}}{(\alpha-1)} t_1^{3-\alpha} h(\theta) \hspace{.8cm} \mathrm{with}\nonumber \\
&&h(\theta) = \alpha 
+ \left(1+\theta\right)^{3-\alpha} + (\alpha-3)\left(1+\theta\right)^{2-\alpha} -\theta^{3-\alpha} ,
\nonumber
\end{eqnarray}
where $\theta=\Delta/t_1$. 
Inserting this expression in Eq. (\ref{eq04}) 
we find 
\begin{equation}
\langle \overline{\delta^2} \rangle = { \langle x^2 \rangle \over \alpha -1} .
\label{taDel}
\end{equation}
Thus, except for the normal diffusion limit of
 $\alpha \rightarrow 2$ 
the ensemble average MSD differs from the time average MSD by a factor.
Numerical evidence for a difference between $\langle x^2\rangle$ and 
$\langle\overline{\delta^2}\rangle$ was presented earlier \cite{Akimoto12} 
in the context of diffusion generated by deterministic maps.
Eq. (\ref{taDel})  quantifies this deviation 
with $\alpha = 1/(z-1)$ and $3/2<z<2$ being the nonlinearity parameter of 
the deterministic map in \cite{Akimoto12}.

To explain this effect note that $\langle x^2 \rangle$ is calculated
for a process which starts at time $t=0$. Physically
this corresponds to a particle immersed in a system at time $t=0$ 
when the process begins. Alternatively we may measure or calculate the
\textit{stationary} MSD $\langle x^2 \rangle_{st}$. This is the MSD
of a process which started long before the measurement  begins at
$t=0$.  In this case, since $\langle \tau \rangle$ is finite, 
the system is in the stationary state throughout the measurement time, i.e. from $t=0$ on. It follows that
one may use the Green-Kubo formalism to obtain 
the stationary MSD
%
$\left\langle x^2 \right\rangle_{st} = 
2 K_{\alpha}t^{3-\alpha}/(\alpha-1) $,
as was done earlier e.g. in \cite{ZumKlaf93a,BarFleu97}.
Hence we find that $\langle \overline{\delta^2} \rangle =\langle x^2 \rangle_{st}\neq \langle x^2 \rangle$. 
The assessment of the ergodic properties of the process in the sense of equal time- and ensemble 
averaged MSD is therefore a subtle issue which depends on the initial preparation of the system.
Such a behavior is not found for normal diffusion processes.

In biophysical experiments the time averaged MSDs of trajectories measured up to a certain 
observation time are often distributed \cite{Caspi02,Bruno09}.
In single particle tracking experiments in the living cell
the origins of these fluctuations and of the anomalous transport are still an 
object of controversy \cite{P2day,osh12,Lau03}. 
Time averages are made over finite times
which are limited by biological function, e.g. 
the measurement time cannot be larger than 
the life time of the cell. Hence the usual infinite long time limit and ideas on stationarity are 
not relevant in many single bio-molecule experiments.
In our simulations we find large fluctuations intrinsic to the L\'evy walk model for 
$\alpha=5/4$ (see left panel of Fig. \ref{fig1}) 
among the $\overline{\delta^2}$ even for $\Delta/T = 10^{-3}$.
To check whether $\overline{\delta^2}$ remains random,
we investigated the fluctuations which vanish in the long time
limit, though slowly (in contrast to L\'evy flights, see \cite{weron}). 
Thus  we find that 
\begin{equation}
\lim_{T\to\infty} {\overline{\delta^2}\over\langle x^2\rangle} = {1\over |1-\alpha|}, \label{eq17}
\end{equation}
both for the ballistic and enhanced diffusion phase. 


\textit{ Response to bias and generalized Einstein relation.}
Now we  assume a small constant force $F$ acting on the particle, 
a case that leads to an anomalous drift. Thereby the force accelerates
the particle according to Newton's law of motion, similarly to the Drude model, 
while the waiting times for the collisions are still drawn from $\psi(\tau)$.
We consider the time average 
Eq. (\ref{EATAdrift})
%
%
and find using $\langle x \rangle_F$  
\begin{eqnarray}
\langle\overline{\delta}\rangle_F &=& 
\left\{
\begin{array}{c l}
(1-\alpha)F T \Delta/(2M) & 0<\alpha<1   \\
K_{\alpha} F T^{2-\alpha} \Delta /(Mv_0^2) & 1<\alpha<2 .
\end{array}
\right.
\label{Fdrift}
\end{eqnarray}
These results  differ  
from their corresponding ensemble average $\langle x\rangle_F$ 
in that they depend on both the lag time  $\Delta$ and the measurement
time $T$. In contrast $\langle x \rangle_F$ clearly depends only
on the measurement time, namely $\langle x \rangle_F = F(1-\alpha)/(2M) t^2$ for $0<\alpha<1$ 
and $\langle x \rangle_F = FK_\alpha/(Mv_0^2) t^{3-\alpha}$ for $1<\alpha<2$ \cite{BarFleu98}. 
The equivalence of time and ensemble averaging is thus
broken. This is a consequence of the of the non-linear dependence of $\langle x\rangle_F$ on time
and thus in fact a very general behavior valid for any system whose response to the driving force 
is anomalous.
The limit of normal diffusion $\alpha \to 2$ renders $\langle\overline{\delta}\rangle_F$ a function
of the lag time only so that ergodicity is retained. 

Using Eqs. (\ref{eq08}), (\ref{taDel}) and (\ref{Fdrift}) we find  the generalized Einstein relation for the time averages
\begin{eqnarray}
\frac{\langle \overline{\delta} \rangle_F}{\langle \overline{\delta^2} \rangle} 
&=& \frac{|1-\alpha|F}{2 {\cal T}_{eff}} \left(\frac{T}{\Delta}\right)^{\gamma}. \label{genEinst}
\end{eqnarray}
Here $\gamma=1$ in the ballistic phase, while in the enhanced phase $\gamma=2-\alpha$. 
The effective temperature is defined with the averaged
kinetic energy of the particle  ${\cal T}_{eff}/2 = M v_0^2/2$ $(k_B=1)$.
Our relation Eq. (\ref{genEinst})
is very different from the standard  Einstein relation
for the ensemble averages
$\langle x \rangle_F / \langle x^2 \rangle = F / (2 {\cal T}_{eff})$ \cite{Bouchaud},
the normal diffusion limit $\alpha \to 2$ being the exception.

\textit{ Conclusion.}
Since the days of Einstein huge attention has been given to ensemble averaged
response functions (e.g. mobility) and its relation to fluctuations
via fluctuation dissipation relations. 
In this letter we have shown that for a widely applicable
class of anomalous processes
the time averages do not obey simple Einstein relations,
contrary to the ensemble averages. 
Due to the lack of a time
scale in the dynamics we find a new type of Einstein relations 
Eq. (\ref{genEinst}) 
which depends on the measurement time and thus exhibits aging.
This new type  of Einstein relations entails a mobility effectively increasing with the measurement time, 
reflecting the large excursions in the L\'evy walk.
Further we have unraveled the nature of the fluctuations of the time averaged MSDs
of the L\'evy walk which exhibit Mittag-Leffler universality 
in the ballistic phase Eqs. (\ref{eq11}), (\ref{eq12}).
In the enhanced phase the fluctuations were comparably large 
though slowly decaying and 
revealed a delicate sensitivity to the initial preparation of the system, 
characteristics that cannot be found for normal processes.

\textit{This work was supported by the Israel Science Foundation. }


\begin{thebibliography}{99}
\bibitem{remark1} To see this denote $t_s$ as the time where the
particle switched
from $+ v_0$ to $-v_0$. Then 
$[x(t'+\Delta) - x(t')]^2 = (v_0 \Delta)^2$ if $t'\le t_s - \Delta$ or
$t_s \le t'$, else it is $(2 v_o t_s - v_0 \Delta - 2 v_0 t')^2$. Inserting
this expression in Eq. (\ref{eq01}) gives $\overline{\delta^2} = (v_0 \Delta)^2 - (2/3) (v_0)^2 \Delta^3 / T$ for $T\gg\Delta$.
\bibitem{remark6}
Provided
the two transitions do not fall within a time interval  $\Delta$.
\bibitem{remark2}  
Here we assume that $1 \ll \Delta$ 
where the
 time scale $1$ refers to the cutoff of 
$\psi(\tau)=\alpha \tau^{-(1+\alpha) } $ for $\tau >1$. 
In the opposite limit
$\Delta \ll 1 \ll T$ we have $\chi^2 =  2v_0^2\Delta^3/(3 T)$.
\bibitem{remark7}
$g(\xi)$ is the density of the so-called Mittag-Leffler distribution.
\bibitem{remark3} We obtain the velocity correlation function Eq. (\ref{eq05a}) in the scaling limit and used Eq. (\ref{eq05})
to find $\langle x(t_1)x(t_1+\Delta)\rangle$ (details will be published elsewhere). To get Eq. (\ref{taDel}) we used the
$\langle x(t_1)x(t_1+\Delta)\rangle$, Eq. (\ref{eq02}) and Eq. (\ref{eq04}).
\end{thebibliography}
\end{document}